\documentclass{article}
\usepackage{spconf,amsmath,graphicx,hyperref,footmisc}

\title{Knowledge Transfer from Weakly Labeled Audio using Convolutional Neural Network for Sound Events and Scenes \vspace{-0.15in}}
\twoauthors
  {Anurag Kumar\vspace{-0.15in}}
	{Carnegie Mellon University, Pittsburgh, PA, USA\\ \vspace{-0.05in} {\small \tt alnu@andrew.cmu.edu} \vspace{-0.25in}}
  {Maksim Khadkevich, Christian F\"{u}gen\vspace{-0.15in}}
	{Facebook Inc. Menlo Park, CA, USA \\
	\vspace{-0.05in}
	{\small \tt khadkevich,fuegen@fb.com} \vspace{-0.25in}}

\begin{document}
\ninept
\maketitle

\begin{abstract}
\vspace{-0.05in}
In this work we propose approaches to effectively transfer knowledge from weakly labeled web audio data. We first describe a convolutional neural network (CNN) based framework for sound event detection and classification using weakly labeled audio data. Our model trains efficiently from audios of variable lengths; hence, it is well suited for transfer learning. We then propose methods to learn representations using this model which can be effectively used for solving the target task. We study both transductive and inductive transfer learning tasks, showing the effectiveness of our methods for both domain and task adaptation.  We show that the learned representations using the proposed CNN model generalizes well enough to reach \emph{human level accuracy} on ESC-50 sound events dataset and sets \emph{state of art} results on this dataset. We further use them for acoustic scene classification task and once again show that our proposed approaches suit well for this task as well. We also show that our methods are helpful in capturing semantic meanings and relations as well. Moreover, in this process we also set state-of-art results on \emph{Audioset} dataset using \emph{balanced} training set. 
\end{abstract}
\vspace{-0.1in}
\section{Introduction}
\vspace{-0.1in}
\label{sec:intro}
Sound plays a crucial role in our interaction with the surroundings. Hence, it is critical for the success of artificial intelligence that machines or computers are able to comprehend sounds as humans do. This has led to considerable interests in sound event detection and classification research in recent years. The motivation also comes from immediate applications such as surveillance \cite{atrey2006audio}, content based indexing and retrieval of multimedia \cite{tong2014lamp,yu2014informedia} to name a few. 

Sound event detection and classification has been constrained by the availability of large scale datasets. Labeling sound events in an audio recording is an extremely difficult task. Besides this there are also situations where marking beginnings and ends of a sound event in an audio recording is inherently ambiguous and open for interpretation by the annotator \cite{kumar2017deep}. To address these concerns \cite{kumar2016audio} introduced weak labeling approaches for sound event detection. Recently, a large scale weakly labeled dataset for sound events, \emph{Audioset} \cite{gemmeke2017audio},  has been released. Weak label learning for sound events was also included in this year's DCASE2017 challenge\cite{mesaros2017dcase}. 

Although weak labeling addresses data availability constraints to a certain extent, creating large datasets along the lines of \emph{Audioset} is still not easy. Even weak labeling, when done manually can be a resource intensive and time consuming process.  Moreover, it might just be difficult to collect large amounts of labeled data in any form in certain cases. For examples, there are sound events which are inherently rare. Deep learning methods such as those based on CNNs is not directly useful in such cases. However, as pointed out by Ellis \emph{et. al.} in Future Perspectives \cite{ellis2018future}, one can attempt to address this problem by transferring knowledge from a model trained on a large dataset. Motivation also comes from computer vision where deep CNN models have been successfully used to transfer knowledge from one domain to another as well as from one task to another \cite{oquab2014learning, yosinski2014transferable}. This approach, more generally referred to as \emph{transfer learning} \cite{pan2010survey} remains more or less unexplored in the context of sound events and scenes. Some audio related works in transfer learning are \cite{deng2013sparse, coutinho2014transfer, dimenttransfer}. In another earlier work \cite{kumar2014detecting}, models are first trained on one set of sound events and then tested on another set to understand the idea of \emph{objectness} in sounds.  

\begin{figure*}[t]
   \centering
   \includegraphics[width=0.98\linewidth]{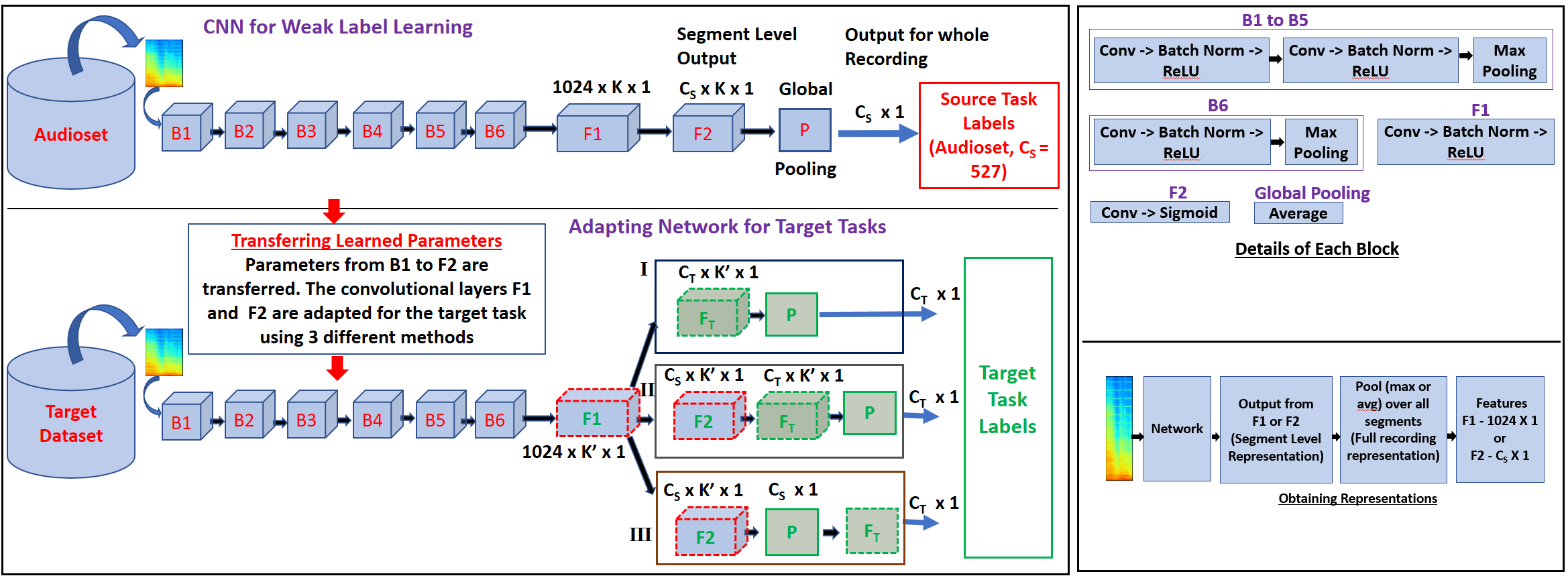}
   \vspace{-0.10in}
     \caption{\textbf{\emph{Top Left and Right:}} Deep CNN for Weakly Labeled Audio. B1 to B6 consists of convolutional and pooling layers. F1 and F2 are convolutional layers. P is global pooling layer which transforms segment level output to recording level output. \textbf{\emph{Bottom Left:}} Adapting CNN for target task. 3 different methods (I, II, III) are proposed.  Parameters from B1 to F1 (or up to F2) are transferred. F1 and (or) F2 onwards are adapted for target task. Newly added layers are shown in green outline. Layers which are updated during task adaptive training are shown in dashed outline. \textbf{\emph{Bottom Right}}: Obtaining representations for audios. Network can be $\mathcal{N}_S$ or one of $\mathcal{N}_T^I$, $\mathcal{N}_T^{II}$, $\mathcal{N}_T^{III}$. See Section \ref{ssec:adapt}.}
     \label{fig:framework}
\vspace{-0.2in}
\end{figure*}

Transfer learning in computer vision has been successful because of the availability of large datasets such as \emph{Imagenet}, which provides a reasonable collection of labeled examples for a large number of visual objects. This allows one to train deep models which can learn enough information from the source data to be useful in solving other tasks. For sounds, the primary problem has been lack of such large scale dataset; by \emph{large scale} we imply both, the vocabulary of sound events as well as the overall dataset size. The vocabulary of sound events is important because a learning algorithm needs to see a wide variety of sound events to learn models which might be useful in solving other tasks. 

Due to lack of such large dataset, Soundnet \cite{aytar2016soundnet} proposed to transfer knowledge from visual models for sound event recognition. They use CNN models trained for visual objects and scenes to teach a feature extractor network for audio. However, it remains to be seen how a more direct approach of audio to audio knowledge transfer can be done. 

In this paper, we propose methods to effectively transfer knowledge from a CNN based sound event model trained on a large dataset (Audioset). We first train a deep CNN model on \emph{Audioset}, a dataset which provides weakly labeled audio examples from \emph{YouTube} for $527$ sound events. Our proposed CNN for weak label learning works efficiently and smoothly with audio recordings of variable length. This makes it well suited for its application transfer learning. Moreover, it also outperforms previous methods \cite{hershey2017cnn} and is computationally much more efficient. 

We then use the above CNN model in both \emph{transductive} as well as \emph{inductive transfer learning} scenarios \cite{pan2010survey}. For transductive learning scenario which might also be referred to as domain adaptation, we use the CNN models for sound event classification on ESC-50 dataset \cite{piczak2015esc}. In this case the target task is still sound event classification, but the audio recordings are from a different domain. For the inductive transfer learning which might also be referred to as task adaptation, we perform acoustic scene classification task on DCASE 2016 dataset \cite{mesaros2016tut}. In  \emph{task adaptation} the target task is different from source task. In both cases, the CNN model serves as the framework to learn representations for audios which can further be used to train a classifier. We propose different methods to adapt the network for the target tasks to obtain discriminative representations. 

Moreover, we show that these representations capture higher level semantic information as well. Our method also helps automatically understand the relationship between acoustic scenes and sound events. To the best of our knowledge, this is the first work which extensively explores and proposes methods to transfer knowledge from a CNN based model trained on a large-scale sound event dataset such as Audioset.


\vspace{-0.1in}
\section{Deep CNN for Weakly Labeled Audio}
\vspace{-0.1in}
\label{sec:weakmodel}
Several CNN approaches have been proposed for sound event classification (SEC), \cite{piczak2015environmental, salamon2017deep} to cite a few. However, most of these works are formulated around \emph{strongly labeled} data. When done on weakly labeled data they are almost always limited in terms of their scale \cite{xu2017attention, su2017weakly, kumar2017deep}; offering little insight into how well they might generalize in large scale scenarios and be useful for transfer learning. The DCASE 2017 \cite{mesaros2017dcase} weakly labeled challenge and works based on it also considers only 17 events from Audioset. 

\cite{hershey2017cnn} analyzes popular CNN architectures such as VGG, Resnet for large scale sound event classification on web videos. However, the training procedure in \cite{hershey2017cnn} makes a simplistic strong label assumption for weakly labeled audios. The sound event is assumed to be present in the whole audio recording. For training CNNs, an audio recording is chunked into small segments and fed one by one to the network and the target labels for all segments is set to be same as the label for the whole recording. This training procedure will be referred to as \emph{strong label assumption training} (SLAT). 

In this work, we give an alternate approach premised on the ideas proposed in \cite{kumar2017deep}, which treats weak labels as weak while training CNNs. Before going into other details, we would like to mention that Logmel spectrograms are used for training CNNs in this work. All audio recordings are sampled at $44.1$ KHz sampling frequency. 128 mel bands are used. A window size of 23 ms and an overlap of 11.5 ms is used for obtainining mel features.  

\vspace{-0.15in}
\subsection{Network Architecture}
\vspace{-0.1in}
Our proposed deep CNN framework for weakly labeled audio is shown at the top left and right panels in Figure \ref{fig:framework}. Block B1 to B5 consists of two convolutional layers (with batch normalization) followed by a max pooling. B6 consists of one colvolutional layer, followed by a max pooling layer. ReLU ($max(0,x)$)\cite{nair2010rectified} activation is used in all cases. For convolutional layers in all six blocks,  $3 \times 3$ filters are used. Stride and padding are fixed to $1$.  The number of filters used in convolutional layer(s) of blocks B1 to B6 are, $\{B1: 16, B2:32, B3:64, B4:128, B5:256, B6:512\}$. Max pooling are done over a $2 \times 2$ window, with a stride of $2$ by $2$. 

F1 is also a convolutional layer with ReLU activation. $1024$ filters of size $2 \times 2 $ are used with a stride of $1$. No padding is used in F1. F2 is the secondary output layer, a covolutional layer of $C_S$ filters of size $1 \times 1$ and sigmoid output. This layer produces segment level output ($C_s \times K \times 1$), where $C_s$ is the number of classes (in source task) and $K$ is the number of segments. The segment level outputs are aggregated using a \emph{global pooling} layer to produce $C_s \times 1$ dimensional output for the whole recording. 

The network scans through the whole input (Logmels) and produces outputs corresponding to segments of $128$ logmel-frames moving by $64$ frames. For example, an input logmel spectrogram consisting of $896$ logmel-frames, that is  $X \in R^{896 \times 128}$ ($128$ mel-bands as stated before),  will produce $K$=$13$ segments at $F1$ and $F2$. Since in weakly labeled  audio we have labels for the full recording, the outputs at segment level are pooled to obtain the full recording level output. The loss is then computed with respect to this recording level output. Hence, this network ($\mathcal{N}_S$) treats weak labels as weak. Overall, it is a VGG style \cite{simonyan2014very} CNN for weak label learning of sounds. The segment size and segment hop size can be controlled by the network design. For $\mathcal{N}_S$, these are $128$ ($\sim$ 1.5 s) and $64$ ($\sim$ 0.75 s) frames respectively in logmel spectorgrams. Note that, if needed segment level outputs can give temporal localization of events \footref{webpg}. 


Unlike SLAT where fully connected dense layers are used in CNN architectures, $\mathcal{N}_S$ network is fully convolutional which allows it to process audio recordings of variable length. This makes it well suited for transfer learning.   
\vspace{-0.1in}
\subsection{Multi-Label Training Loss}
\label{ssec:mltloss}
\vspace{-0.1in}
For web data, including Audioset, it is expected that multiple sound events might be simultaneously present in the same recording. Hence, we employ a multi-label training loss. The sigmoid output gives class specific posteriors for any given input. The binary cross entropy loss with respect to each class is given by $l(y_c,p_c) = -y_c*log(p_c) - (1-y_c)*log(1-p_c)$. $y_c$ and $p_c = \mathcal{N_S(X)}$ are target and network output for $c^{th}$ class respectively. The training loss is the mean of losses over all classes, $L(\mathcal{X},y) = \frac{1}{C} \sum_{c=1}^C l(y_c,p_c)$. 

\vspace{-0.15in}
\section{Transfer and Representation Learning }
\vspace{-0.1in}
\label{sec:transferlrng}
The network $\mathcal{N}_S$ trained on source task audios is used to obtain representations for audios in the target task. The flow of obtaining representations is shown in the bottom right panel of Fig \ref{fig:framework}. Segment level outputs from F1 ($1024 \times K \times 1$) and F2 ($C_S \times K \times 1$) serves as base representations for audios. These segments level representations are then mapped to full recording level representations. We apply either $max()$ or $avg()$ for this mapping. Finally, we obtain $1024$ (F1) or $C_S$ (F2) dimensional representations for full recordings. 


During $\mathcal{N}_S$ training, the blocks from B1 to B6 embeds knowledge from source audio data into $F1$, which is then mapped to source labels by filters in $F2$. This makes F1 well suited for transfer learning, where it can be used to train classifiers for target task. Moreover, outputs from $F2$ gives us a distribution over the source labels, which itself can be useful for the target task when $\mathcal{N}_S$ is trained over a large collection of sound events. We propose two broad methods for representation learning for audios in target tasks using $\mathcal{N}_S$.  
\vspace{-0.15in}
\subsection{Direct Off-the-shelf Representations}
\vspace{-0.1in}
\label{ssec:subh}
In this method, $\mathcal{N}_S$ is treated in a ready to use mode for obtaining representations. Logmel spectrograms of audio recordings from target task are fed into $\mathcal{N}_S$ and the outputs from F1 and F2 are aggregated over all segments (as described before) to obtain $1024$ and $C_S$ dimensional representations respectively. 
\vspace{-0.15in}
\subsection{Transfer and Adapt for Learning Representations}
\label{ssec:adapt}
\vspace{-0.1in}
In the second method, we first adapt the network to target task to extract features which we expect will be more discriminative and better suited for the target classification task. We propose 3 methods to achieve this goal. The methods are shown in Figure \ref{fig:framework}. In all three methods, the parameters from B1 to B6 are transferred and are not updated during the target adaptation training.  Let $C_T$ be the number of classes in the target dataset. 

\textbf{Method I ($\mathbf{\mathcal{N}_T^{I}}$) : } $\mathcal{N}_T^{I}$ performs a direct adaptation of F1 to the target task. Here, F2 is replaced by a new covolutional layer ($F_T$) of $C_T$ filters. Parameters in F1 and $F_T$ are then updated using the training set of the target task. We will call this network $\mathcal{N}_T^I$.  

\textbf{Method II ($\mathbf{\mathcal{N}_T^{II}}$) : } In $\mathcal{N}_T^{II}$, a new convolutional layer ($F_T$) of with $C_T$ filters is added after F2 in $\mathcal{N}_S$. This new network, $\mathcal{N}_T^{II}$, is then adapted for the target task. As shown by dashed boundaries in Fig \ref{fig:framework}, during the adaptive training only F1, F2 and $F_T$ are updated. The idea is to capture target specific information by first transitioning to source label space (F2) and from there going to target label space. 

\textbf{Method III ($\mathbf{\mathcal{N}_T^{III}}$) : } In $\mathcal{N}_T^{III}$, a new fully connected layer $F_T$ of size $C_T$ is added after the global pooling layer in $\mathcal{N}_S$. Once again, only F1, F2 and $F_T$ are updated during network adaptation training. The motivation behind this network is same as $\mathcal{N}_T^{II}$, except that it tries to learn the mapping at full recording level instead of segment level. Note that in both $\mathcal{N}_T^{II}$ and $\mathcal{N}_T^{III}$, the activation function in F2 is changed to ReLU from sigmoid in $\mathcal{N}_S$. 

For all three adapted networks $\mathcal{N}_T^I$, $\mathcal{N}_T^{II}$ and $\mathcal{N}_T^{III}$, if the target task is a multi-label problem, then the activation in final layer is kept as sigmoid and loss function similar to that defined in Section \ref{ssec:mltloss}  is used. However, if the target task audios have single label, then we can use \emph{softmax} output with categorical cross entropy loss. 

A few things worth noting. First, the target task can have audio recordings of different length and our proposed methods can handle such cases efficiently. Moreover, the target task dataset can either be strongly or weakly labeled and the proposed methods can be used to learn representations in both cases.  Lastly, to emphasize again,  the focus is on exploiting $\mathcal{N}_S$ to learn representations for audios in the target task. Classifier such as SVM can be trained on these representations, even if the target task dataset is small. 

\begin{table}[t]
   \centering
\resizebox{0.53\columnwidth}{!}{
\begin{tabular}{|c|c|c|c|}
	\hline  
\multicolumn{2}{|c|}{\textbf{MAUC}} &  \multicolumn{2}{c|}{\textbf{MAP}}\\ 
	\hline
	  $\mathbf{\mathcal{N}_S^{slat}}$ & $\mathbf{\mathcal{N}_S}$ & $\mathbf{\mathcal{N}_S^{slat}}$ & $\mathbf{\mathcal{N}_S}$\\
	\hline
	0.915 & 0.927 (\textbf{+1.3\%}) &0.167&0.213 (\textbf{+27.5\%}) \\
	\hline
\end{tabular}
}   
\resizebox{0.45\columnwidth}{!}{
\begin{tabular}{|c|c|c|c|}
	\hline  
\multicolumn{2}{|c|}{\textbf{Train Time}} &  \multicolumn{2}{c|}{\textbf{Inference Time}}\\ 
	\hline
	  $\mathbf{\mathcal{N}_S^{slat}}$ & $\mathbf{\mathcal{N}_S}$ & $\mathbf{\mathcal{N}_S^{slat}}$ & $\mathbf{\mathcal{N}_S}$\\
	\hline
	1.0 & 0.61 (-39\%) & 1.0 & 0.67 (-33\%)\\
	\hline
\end{tabular}
}  
  
\resizebox{0.99\columnwidth}{!}{   
\begin{tabular}{|c|c|c|c|c|c|}
	\hline  
\multicolumn{3}{|c|}{\textbf{Lowest 10}} &  \multicolumn{3}{c|}{\textbf{Highest 10}}\\ 
\hline
	Event & $\mathcal{N}_S^{slat}$ & $\mathcal{N}_S$ & Event & $\mathcal{N}_S^{slat}$ & $\mathcal{N}_S$\\
	\hline
	Scrape & 0.0058&0.0092   & Music&0.728&0.749 \\
	\hline
	Crackle & 0.0078&0.0097   &Siren (Civil Defense)&0.671&0.641 \\
	\hline
	Man Speaking & 0.0080&0.0202  &Bagpipes&0.646&0.786 \\
	\hline
	Mouse & 0.0092&0.0368    &Speech&0.631&0.661 \\
	\hline
	Buzz & 0.0095&0.0077    &Purr (Cats)&0.575&0.600 \\
	\hline
	Squish & 0.0102&0.0122   &BattleCry&0.575&0.651 \\
	\hline
	Gurgling & 0.0111&0.0125   &Heartbeat&0.559&0.569 \\
	\hline
	Door & 0.0115&0.0685   &Harpsichord&0.544&0.630 \\
	\hline
	Noise & 0.0116&0.0107    &Ringing (Campanology)&0.538&0.690 \\
	\hline
	Zipper & 0.0121&0.0161    &Timpani&0.538&0.528 \\
	\hline
	\textbf{Mean} & \textbf{0.0097}&\textbf{0.0203}& \textbf{Mean} &\textbf{0.600}&\textbf{0.651} \\
	\hline
\end{tabular}
}
\vspace{-0.1in}
\caption{\textbf{\emph{Top Left}}: Comparison of MAUC and MAP over all $527$ events in Audioset. \textbf{\emph{Top Right}}: Comparison of Average Relative Training ( 1 Epoch) and Inference (per test instance) times. \textbf{\emph{Bottom Table}}: AP comparison for 10 sound events with lowest and highest APs using baseline $\mathcal{N}_S^{slat}$. Section \ref{ssec:audres} and here\textsuperscript{3} for details.}    
\vspace{-0.25in}
\label{fig:audiosetres}
\end{table}

\vspace{-0.2in}
\section{Experiments and Results}
\label{sec:expt}
\vspace{-0.15in}
We start by showing performance for sound event classification on \emph{Audioset}\footnote{https://research.google.com/audioset/,  \textsuperscript{2}http://pytorch.org/}. We work with all $527$ sound events in Audioset for which weakly labeled data is currently available. We compare performance of our $\mathcal{N}_S$ with SLAT ($\mathcal{N}_S^{slat}$). $\mathcal{N}_S^{slat}$ is similar to $\mathcal{N}_S$ except that F1 and F2 are now fully connected layers of size $1024$ and $C_S$. Training is done with fixed size segments of 128 logmel frames as inputs, segments overlap by 64 frames. Loss is computed for each input segment by using recording level labels. 

All experiments are done in Pytorch\textsuperscript{2}. Adam optimization \cite{kingma2014adam} is used for training networks. Validation set is used for tuning parameters and selecting the best model. 

We then show experimental results for transfer learning using $\mathcal{N}_S$. For the task adaptive training of $\mathcal{N}_S$, the training set of the target task is used. Learning rate during this process is fixed to $0.0002$ and updates are done for $50$ epochs, after which the network is used to obtain representations. Linear SVMs \cite{fanliblinear} are then trained on the representations obtained from different methods. The slack parameter $C$ in SVMs is tuned by cross validation on the training set. 

Due to \emph{space constraints} readers are requested to visit \textbf{this webpage} \footnote{\label{webpg}\url{http://www.cs.cmu.edu/\%7Ealnu/TLWeak.htm}} for more detailed results and analysis. 
\vspace{-0.20in}
\subsection{Audioset Results}
\vspace{-0.1in}
\label{ssec:audres}
Audioset\textsuperscript{1} dataset consists of weak labels for $527$ sound events on YouTube videos. Total dataset consists of over 2 million audio recordings. We use the \emph{balanced training} set for training $\mathcal{N}_S$. \emph{Balanced} training set provides a total of around $22,000$ training audio recordings, with at least $59$ examples per class. However, due to multi-label nature of the data the actual number of examples for several classes is much higher. A small subset from \emph{Unbalanced} set of Audioset is used as the validation set in experiments. Results  are reported on the full \emph{Eval} set of Audioset, which has around $20,000$ test audio recordings, again with at least $59$ examples per class. Similar to \cite{hershey2017cnn}, Area under ROC curves (AUC) \cite{fawcett2004roc} and Average Precision (AP) \cite{buckley2004retrieval} for each class are used as evaluation metrics. 

Table \ref{fig:audiosetres} shows Mean AUC (MAUC) and Mean AP (MAP) over all $527$ classes in Audioset. An absolute improvement of $1.2$ ( $1.3\%$ relative) in MAUC and $4.6$ ($27.5\%$ relative) in MAP is obtained using $\mathcal{N}_S$. The top right table shows relative computational times, normalized for comparison. $\mathcal{N}_S$ is $33\%$ faster on an average during inference. Hence, more suitable for real applications. During training, on an average it is $39\%$ faster for $1$ full pass over training data. 

Performance of all classes are available here\footref{webpg}. Bottom table in Tables \ref{fig:audiosetres} shows comparison for 10 sound events for which $\mathcal{N}_S^{slat}$ achieved least and highest APs. For low performance classes,  on an average $\mathcal{N}_S$ doubles the AP (\textbf{0.0097} to \textbf{0.0203}). For easier sound classes \textbf{8.5\%} relative improvement is obtained using $\mathcal{N}_S$. 
\begin{table}[t]
\resizebox{0.41\columnwidth}{!}{
\begin{tabular}{|c|c|}
	\hline  
Methods & Mean \\
 & Accuracy\\
\hline
Piczak \cite{piczak2015environmental} & 64.5 \% \\
\hline
Tokozume \cite{tokozume2017learning} & 71.0 \% \\
\hline
Aytar \cite{aytar2016soundnet} & 74.2 \% \\
\hline
\textbf{Proposed} (\textbf{F1}) & \textbf{83.5} \% \\
	\hline
\end{tabular}
}
\resizebox{0.59\columnwidth}{!}{
\begin{tabular}{|c|c|c|c|c|}
	\hline  
Network	&\multicolumn{2}{c|}{\textbf{F1}} &  \multicolumn{2}{c|}{\textbf{F2}}\\ 
	\cline{2-5}
	 & $max()$ & $avg()$ & $max()$ & $avg()$\\
	\hline
	$\mathbf{\mathcal{N}_S}$ & \textbf{82.8} & 81.6&65.5&64.8 \\
	\hline
	$\mathbf{\mathcal{N}_T^{I}}$ & \textbf{83.5}&81.3&--&-- \\
	\hline
	$\mathbf{\mathcal{N}_T^{II}}$ & \textbf{83.5}&81.8&81.9&81.5 \\
	\hline
	$\mathbf{\mathcal{N}_T^{III}}$ & 83.3&82.6&82.6&81.9 \\
	\hline
\end{tabular}
}
\vspace{-0.15in}
\caption{\textbf{\emph{Left}}: ESC-50 Accuracy comparison with baselines. \textbf{\emph{Right}}: Accuracy comparison of different representations. }  
\vspace{-0.1in}
\label{tab:esc50res}
\end{table}
\begin{table}[t]
\resizebox{0.99\columnwidth}{!}{   
\begin{tabular}{|c|c|c|c|c|c|}
\hline
	Scene & Baseline & $\mathcal{N}_S^{III}$ (F1, $max()$) & Scene & Baseline & $\mathcal{N}_S^{III}$ (F1, $max()$)\\
	\hline
	Beach & 69.3 &71.9   & Library & 50.4 &73.6 \\
	\hline
	Bus & 79.6&82.4   &Metro Station& 94.7 &80.2 \\
	\hline
	Cafe & 83.2&73.8  &Office& 98.6&85.1 \\
	\hline
	Car & 87.2&89.9    &Park&13.9& 46.9 \\
	\hline
	City Center & 85.5 & 93.3    &Residential Area&77.7&63.9 \\
	\hline
	Forest Path & 81.0& 97.4   &Train &33.6&52.3 \\
	\hline
	Grocery Store & 65.0& 84.6   &Tram &85.4&84.0 \\
	\hline
	Home & 82.1& 69.4   & \textbf{Mean}& \textbf{72.5}&\textbf{76.6} \\
	\hline
\end{tabular}
}
\resizebox{0.99\columnwidth}{!}{
\begin{tabular}{|c|c|c|c|c|c|c|c|c|c|}
	\hline  
Network	&\multicolumn{2}{c|}{\textbf{F1}} &  \multicolumn{2}{c|}{\textbf{F2}} & Network	&\multicolumn{2}{c|}{\textbf{F1}} &  \multicolumn{2}{c|}{\textbf{F2}}\\ 
\hline
	 & $max()$ & $avg()$ & $max()$ & $avg()$  &  & $max()$ & $avg()$ & $max()$ & $avg()$\\
	\hline
	$\mathbf{\mathcal{N}_S}$ & 72.2 & 69.8& 59.1 & 60.4 & $\mathbf{\mathcal{N}_T^{II}}$ & 75.5&73.0&73.8&73.9 \\
	\hline
	$\mathbf{\mathcal{N}_T^{I}}$ & 75.2&73.7&--&-- & $\mathbf{\mathcal{N}_T^{III}}$ & 76.6&73.7&72.5&73.3 \\
	\hline
\end{tabular}
}
\vspace{-0.1in}
\caption{\textbf{\emph{Upper}}: DCASE 2016 accuracy comparison with baseline \textbf{\emph{Lower}}: Accuracy comparison of different representations.}    
\vspace{-0.2in}
\label{tab:dcaseres}
\end{table}
\vspace{-0.1in}
\subsection{Domain Adaptation: Sound Event Classification}
\vspace{-0.1in}
In this section, $\mathcal{N}_S$ is used for learning representations for ESC-50 \cite{piczak2015esc} dataset. ESC-50 dataset consists of a total of 50 sound events, 10 from each 5 broad categories, \emph{Animals} (e.g Dog), \emph{Natural Soundscapes and Water Sounds} (e.g Chirping Birds), \emph{Human Non Speech Sounds} (e.g Clapping), \emph{Domestic Sounds} (e.g. Clock Alarm) and \emph{Exterior Sounds} (e.g Helicopter). The dataset consists of a total of $2,000$ recordings. It comes pre-divided into $5$ folds. Four folds are used for training and validation and the remaining fold is used for testing. This is done all 5 ways and average accuracy across all 5 runs accuracy is reported. Human accuracy on this dataset is $81.3\%$. 

Left Table in Tab.~\ref{tab:esc50res} compares mean accuracy over all 50 classes with state-of-art on ESC-50 dataset. We outperform the best method by $\mathbf{9.3}\%$, setting state-of-art results on ESC-50. Right table in Tab.~\ref{tab:esc50res} shows performance of different representations proposed in this work. Note that,  even off-the-shelf F1 representations using $\mathcal{N}_S$ is able to achieve better than human performance on this dataset. This shows that $\mathcal{N}_S$ does an excellent job in capturing sound event knowledge. Task adaptive training gives further improvement. $max()$ mapping for converting segment level representations of F1 or F2 to full recording representations performs better. The sigmoid output from F2 in $\mathcal{N}_S$ does not give good performance using linear SVMs. However, after task adaptive training in $\mathcal{N}_T^{II}$ and $\mathcal{N}_T^{III}$,  where F2's activations are changed to ReLU, we obtain good performance from F2 representations. Classwise confusion matrix can be found here\footref{webpg}. 
\vspace{-0.2in}
\subsection{Task Adaptation: Acoustic Scene Classification}
\vspace{-0.1in}
Scenes such as \emph{Park} or \emph{Home} possess complex acoustic characteristics. Often, they are themselves composed of several sound events meshed together in a complex manner. We study the utility of transferring learning for acoustic scenes on DCASE 2016 \cite{mesaros2016tut} dataset. It provides $30$ seconds examples for $15$ acoustic scenes listed in upper table in Tab. \ref{tab:dcaseres}. The total duration of data is around $9.75$ hours. The dataset comes pre-divided into $4$ folds, $3$ are used for training and remaining for test. This is done all $4$ ways are average accuracies across all 4 runs are reported. 

Upper table in Table \ref{tab:dcaseres} compares accuracies for different acoustic scenes between baseline and one of our proposed method. An absolute improvement of $4.1\%$ over all 15 scenes is observed. For certain scenes which are hard to classify such as \emph{Park} and \emph{Train}, an absolute improvement of $33.0\%$ and $18.7\%$ respectively is obtained. Note that for this task, representations from task adapted networks perform much better compared to those obtained directly from $\mathcal{N}_S$. $\mathcal{N}_T^{III}$ gives best results, followed closely by $\mathcal{N}_T^{I}$ and $\mathcal{N}_T^{III}$. Once again $max()$ mapping performs better compared to $avg()$. 

\begin{figure}[t]
   \centering
   \includegraphics[width=0.98\columnwidth]{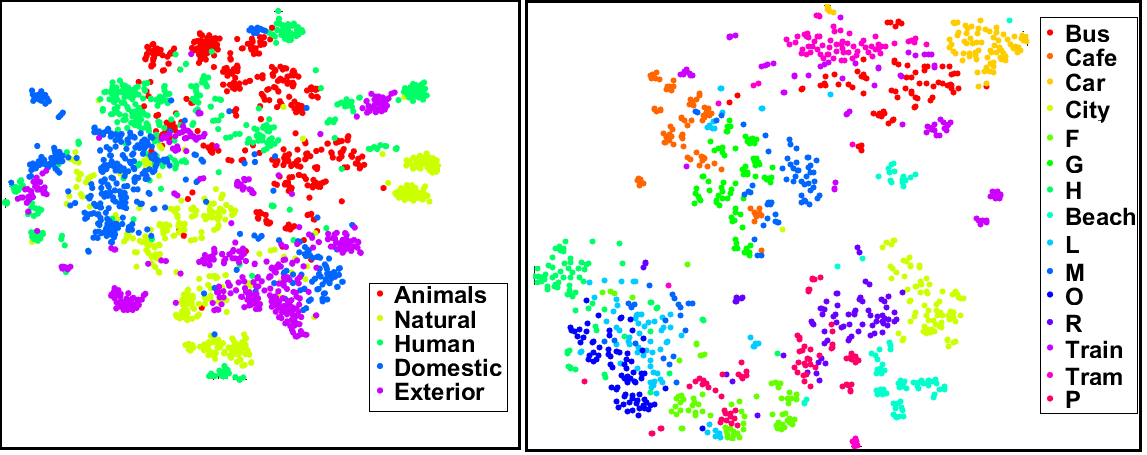} 
\resizebox{0.99\columnwidth}{!}{   
\begin{tabular}{|c|c|}
	\hline  
\textbf{Scene} &  \textbf{Frequent Highly Activated Sound Events}\\ 
\hline
	Cafe & Speech, Chuckle-Chortle, Snicker, Dishes, Television\\
	\hline
	City Center & Applause, Siren, Emergency Vehicle, Ambulance\\
	\hline
	Forest Path & Stream, Boat Water Vehicle, Squish, Clatter, Noise, Pour\\
	\hline
	Grocery Store & Shuffle, Singing, Speech, Music, Siren \\
	\hline
	Home & Speech, Finger Snapping, Scratch, Dishes, Baby Cry, Cutlery\\
	\hline
	Beach & Pour, Stream, Applause, Splash - Splatter, Gush \\
	\hline
	Library & Finger Snapping, Speech, Fart, Snort \\
	\hline
	Metro Station & Speech, Squish, Singing, Siren, Music \\
	\hline
	Office & Finger Snapping, Snort, Cutlery, Speech, Cutlery \\
	\hline
	Residential Area & Applause, Crow, Clatter, Siren \\
	\hline
	Park & Bird Song, Crow, Stream, Wind Noise, Stream \\  
	\hline
\end{tabular}
}
\vspace{-0.1in}
\caption{\textbf{\emph{Top Left}}: t-SNE visualizations for ESC-50 ($\mathcal{N}_S$, F1, $max()$). Color coded for 5 higher semantic categories  \textbf{\emph{Top Right}}: t-SNE visualizations for DCASE 2016. First alphabet for some, e.g (F)orest. \textbf{\emph{Bottom Table}}: Sound Events which are frequently among Top 5 maximally active events for a given scene. Network is $\mathcal{N}_T^{III}$.}    
\vspace{-0.25in}
\label{fig:semantics}
\end{figure}

\vspace{-0.15in}
\subsection{Semantic Understanding}
\vspace{-0.1in}
We now try to draw some semantic inferences from the proposed methods. Left panel in Fig \ref{fig:semantics} shows $2$ dimensional t-SNE \cite{maaten2008visualizing} embeddings for representations obtained for ESC-50. The embeddings are color coded for the $5$ broad categories in ESC-50 dataset, semantically higher level groups for sound events. One can note from the plot that these representations are capable of capturing higher level semantic information. \emph{Vacuum Cleaner} in \emph{Domestic} closely resembles \emph{Chainsaw}, \emph{Engine} and \emph{Handsaw} in \emph{Exterior} category and its representations also  lies  closer to \emph{Exterior} sounds (blue dots among purple). Similarly, visualization for $15$ acoustic scenes is shown in right panel in Fig. \ref{fig:semantics}. 

Acoustic scenes can often be understood through sound events. Each neuron in F2 is essentially representing a sound event class and the activations of these neurons can be used to understand scene-event relations. For each input of a given scene we list the Top 5 maximally activated neurons (events) in F2 (for each segment). We then note the events which occurred most frequently (among top 10) in these lists. These highly active events for some of the scenes are shown in table in Fig \ref{fig:semantics}. We observe that several of these sound events are expected to occur in the corresponding scene. Hence, these scene-events relations are semantically meaningful. This shows that our network managed to successfully transfer knowledge and learn relationships. More analysis on semantics here\footref{webpg}.  
\vspace{-0.20in}
\section{Conclusions}
\label{sec:page}
\vspace{-0.15in}
We first proposed a CNN based model for weakly labeled learning of sound events. Our model not only sets state-of-art results on Audioset but is also computationally efficient and well suited for transfer learning. We then proposed methods to learn representations for audios in the target task using this CNN model. We set state of art results on ESC-50 dataset, achieving an accuracy of $\mathbf{83.5}\%$, which surpasses human accuracy on this dataset. Besides achieving excellent performance, these methods to transfer knowledge are also helpful in higher level semantic understanding. For example, automatically discovering relationships between scenes and sound events is also an important contribution of this work. 


\vfill\pagebreak

\ninept
\bibliographystyle{IEEEbib}
\bibliography{references}

\end{document}